# Effect of microstructure on polymer chain dynamics in polyethylene glycol solutions

Shipra Bhatt [a] and Debjani Bagchi *[a]

*[a]Department of Physics, Faculty of Science, The Maharaja Sayajirao University of Baroda, Vadodara, Gujarat 390002, India*
*Email: debjani.bagchi@gmail.com*

Effect of microstructures and interactions on segmental dynamics in polyethylene glycol (PEG) solution in water is probed with macro-scale oscillatory rheology and micro-scale diffusion of a fluorescent probe. PEG solution fluorescence recovery after photobleaching (FRAP) curves have immobile fractions which increase with PEG concentration, for PEG volume fraction (c) > 0.2, indicating structuring. PEG solution micro-scale diffusion coefficients follow Rouse scaling $D \sim c^{-0.54}$ for c < 0.8 (c*=0.03), resembling unentangled neutral polymers in good solvent. Small amount (0.01 - 1 wt%) nanoclay bentonite (B) in PEG matrix slows down probe diffusion 3-7 times, with heterogeneous dynamics. With 0.01 – 1 wt% carboxymethyl cellulose (CMC) in PEG matrix, probe diffusion is homogeneous with ~10% enhancement in diffusion time. The macroscale storage modulii (G') for PEG, PEG+B, and PEG + CMC solutions scale as viscous fluid-like power law $\omega^\alpha$ with α=2 for $\omega < \tau_d^{-1}$, with $\tau_d$ the terminal relaxation time, followed by short elastic plateau for $\tau_d^{-1} \leq \omega \leq \tau_e^{-1}$. For the regime $\omega > \tau_e^{-1}$, the scaling is a concentration-dependent power law $\omega^\alpha$, with α greater than the Rouse scaling of 0.5 for 0.1<c<0.2 PEG solutions. We identify a time scale $\tau_b$ due to intermolecular interactions in PEG, such that for $\omega > \tau_b^{-1}$, Rouse scaling is recovered. Addition of CMC to PEG restores Rouse scaling. Addition of B gives contributions from both polymer matrix and network of B particles, leading to departure from pure Rouse behaviour. Static microtructural studies reveal clay aggregation due to depletion interactions on increasing the concentration of clay particles in PEG matrix, which leads to a non-monotonic concentration dependence of G' with c at 1 wt% B.

## 1. Introduction

Polyethylene glycol (PEG) is a widely used biodegradable polymer for several biomedical applications such as drug delivery, nanomedicine, biodegradable scaffolds for wound healing or 3D bioprinting of artificial tissues, each requiring tunability of mechanical properties of the PEG matrix[1–5]. For this purpose, fillers such as clay particles, semiflexible cellulose fibres, carboxymethyl cellulose (CMC) polyelectrolyte are added with cross-linking of the PEG matrix in order to achieve the desired mechanical modifications in PEG with minimal cellular toxicity[6,7]. Moreover, target specific drug release necessitates penetration through phospholipid membrane barriers, flow through intracellular and extracellular matrix, or blood plasma, all of which are viscoelastic media. As a result, structural integrity and control of drug release is affected by the forces these nano-medicine polymer complexes are subject to in the viscoelastic media. Hence, in addition to the static mechanical properties, a thorough study of the dynamical mechanical response to external perturbations is imperative for the polymer-filler matrices to fulfil their desired role.

PEG molecules have dynamic intra molecular hydrogen bonding (H-bonds) as well as H-bonds with water molecules. NMR experiments have revealed that water H-bonds bridge neighbouring chains leading to reversible inter-molecular associations[8]. The transient association-dissociation kinetics as well as topological constraints introduced due to H-bonds can complicate the segmental and chain relaxation dynamics. Neutron spin echo experiments on hydrogen bonding polymers have revealed heterogeneities in dynamics, with departure from Rouse dynamics at length scales larger than few Kuhn lengths due to intermolecular and intramolecular associations[9]. Experiments on associative polymers have shown that segmental relaxation in concentrated solutions and melts, which is a faster process compared to interaction dynamics, can be well separated in frequency sweep oscillatory shear experiments, with a higher value of scaling exponent than that of Rouse dynamics (0.5) preceding the Rouse relaxation, permitting an estimation of the activation energy involved in sticky intermolecular interactions[10]. To explain the complicated mechanical response of these systems, theoretical models have incorporated reversible network formation into the classical Rouse or reptation models, called 'sticky Rouse' or 'sticky reptation' model[11–15]. Although simulations have probed the effect on statics, dynamics and rheology of these reversible networks in associating systems, they are limited due to system size[16–18]. Moreover, number, strength and lifetime of these sticky bonds and their influence on stress relaxation is system specific, so for curating PEG-based applications with mechanical tunability, it is important to first study relaxation dynamics under various conditions.

Incorporation of fillers such as clay particles in the PEG matrix to improve mechanical properties introduces additional interactions such as depletion interactions and clay-polymer associative networks, which modulate stress relaxation[19–24]. In addition, impediments in the routes to tunable conditions in polymer-filler composites such as concentration and flow-induced shear banding, yielding, gelation, ageing and micro structuring can arise, necessitating simultaneous measurement of solution microstructure[19,25–29]. Scattering studies under flow and at rest reveal that structural changes are crucial in understanding mechanical response[19,21,25,30,31]. Several experimental, theoretical and simulation studies on nanoparticle-reinforced polymer matrices have revealed a varying degree of mechanical modification of the host polymer matrix by nanoparticles as a result of additional length and time scales introduced due to interactions[16,32–37].

Hence development of PEG-based applications requiring mechanical tunability warrants a thorough investigation of mechanical properties both at the micro scale and the macro scale, especially in the context of modifications due to the inherent 'sticky' interactions and the concentration and type of filler added, which introduce additional interactions. This leads to another question about whether mechanical tunability can be achieved by simple changes in the concentration of the constituents, which can ease large-scale industrial production. Moreover, it is necessary to check for heterogeneities in the system due to competing interactions, because presence of heterogeneities can affect the tribology of films of these nanocomposites enhancing susceptibility to wear, cavitations, fibrillations and fracture. The aim of the present study is to explore the above aspects experimentally by studying dynamics and response to external perturbations at the macroscopic scale and to uncover the correlations of the macroscopic response with structure and dynamics at the microscopic scale. More specifically, this study also aims to explore the effect of associative intermolecular interactions between polymer chains and phase separation due to depletion interactions in presence of colloidal clay particles, on the relaxation dynamics under oscillatory shear.

Predictably, we find that macroscopic mechanical properties are influenced by solution microstructure. Our studies also show that the mechanical response cannot be tuned easily because of its non-trivial relation with polymer as well as filler concentration and strain amplitude. Several features of macro-scale rheological response of PEG-nanoparticle solutions are well explained by the micro-scale mechanical response of these solutions studied with the help of Fluorescence recovery after photobleaching (FRAP) technique, which also reveals the presence of micro-scale heterogeneities. In the solution state, we do see early onset of yielding in the amplitude response of PEG-bentonite systems at higher bentonite concentrations which show pronounced low $q$ upturn in structure factor, attributed to microstructural heterogeneities and bentonite aggregates revealed by phase contrast imaging. The present study reveals the caveat that one cannot achieve a desired mechanical property just by concentration changes, and for applications which require a high level of homogeneity in the polymer matrix, regulation of concentration of fillers to an optimum value in the polymer matrix is important. We observe that the depletion forces present in PEG-bentonite mixture result in widely different viscoelastic response and flow behaviour compared to CMC-PEG blend, which have predominantly short-ranged electrostatic interactions. Based on our observations, we propose that for modelling the viscoelastic flows of these systems, intermolecular interactions cannot be ignored in the constitutive equations.

## 2. Materials and Methods

### 2.1. Materials

Polyethylene Glycol (PEG) of molecular weight 20,000M was purchased from Alfa Aesar and used without further processing. Aqueous solutions of PEG were prepared by weighing the appropriate amount of PEG for 5, 10, 20 and 50 percent by weight solution concentration, and dissolved in deionized water (LOBA Chemicals) by stirring with a magnetic stirrer for about an hour at room temperature, around 30 °C. For the fluorescence microscopy experiments, fluorescein disodium salt (Alfa Aesar) was added to get a concentration of 30μM fluorescein in PEG solution. Solutions of PEG having 0.01, 0.1 and 1 wt.% of nanoclay bentonite (Sigma) and carboxy methyl cellulose (sodium salt, 90,000M molecular weight, the degree of substitution = 0.7 carboxymethyl groups per anhydroglucose unit, from Sigma) were prepared following the procedure mentioned above. To maintain a uniform effect of ageing of clay solutions on the experimental data, all samples were prepared 1 day prior to doing the experiments. All the solution concentrations used in performing experiments for different samples are given in Table 1 and 2 with a nomenclature which will be used throughout the text. The entanglement (or overlap) concentration in volume fraction is calculated as c* = 0.03 (Table 1), and all concentrations for the samples investigated, given as volume fractions, show that our experiments probe the semi dilute to the concentrated polymer solution regimes.

### 2.2. Oscillatory rheology

Oscillatory rheology experiments were performed with Anton Paar MCR 301 rheometer in plate-plate geometry with plate diameter 50mm and 0.1mm gap between the plates. The experiments were performed in a room maintained at 25 °C. Data analysis was done using Microcal Origin.

### 2.3. Fluorescence recovery after Photobleaching (FRAP)

FRAP experiments were carried out with Olympus IX83 FluoView FV3000 Confocal Scanning Laser Microscope (CSLM) with 60x/1.35 NA Olympus UPLAN APO oil immersion objective. Fluorescence of the fluorescein in the solution was excited with a 488nm source (OBIS Coherent Laser) and emission was collected at 530 nm in the XY scanning mode of the confocal microscope. For all the FRAP experiments, the samples were injected into sealed microchannels having a coverslip at one end, which was in contact with the oil immersion objective. 100% laser intensity was used for photobleaching a 100x100 pixels region (Figure 1A) in an image of size 512x512 pixels (0.414 μm per pixel) for a duration of 2 sec. Time lapse images (~1 sec per frame) were acquired before photobleaching and after photobleaching at around 5% laser intensity. This procedure was repeated for five different regions in the microchannel to get an idea of heterogeneities and for statistical accuracy. All FRAP experiments were conducted in a room maintained at 22 °C. Data analysis was performed using custom made programs in MATLAB.A 512 pixels array line intensity profile was extracted from the center of the bleached spot matrix of the first postbleach image, and a Gaussian fit (Equation 1) to this gave a bleach spot width $r_e$ of about 91.3 ± 5 pixels or 37.82 ± 2 μm (Figure 1B)[38].

$$f(x) = 1 - k\, exp\left(-\frac{2(x-c)^2}{r_e^2}\right) \qquad (1)$$

The total fluorescence intensity of the bleached ROI from prebleach and postbleach images was normalized and corrected for photofading using Equation 2[39,40]:

$$I_N(t) = \frac{I_{data} - I_0}{I_{whole} - I_0} \qquad (2)$$

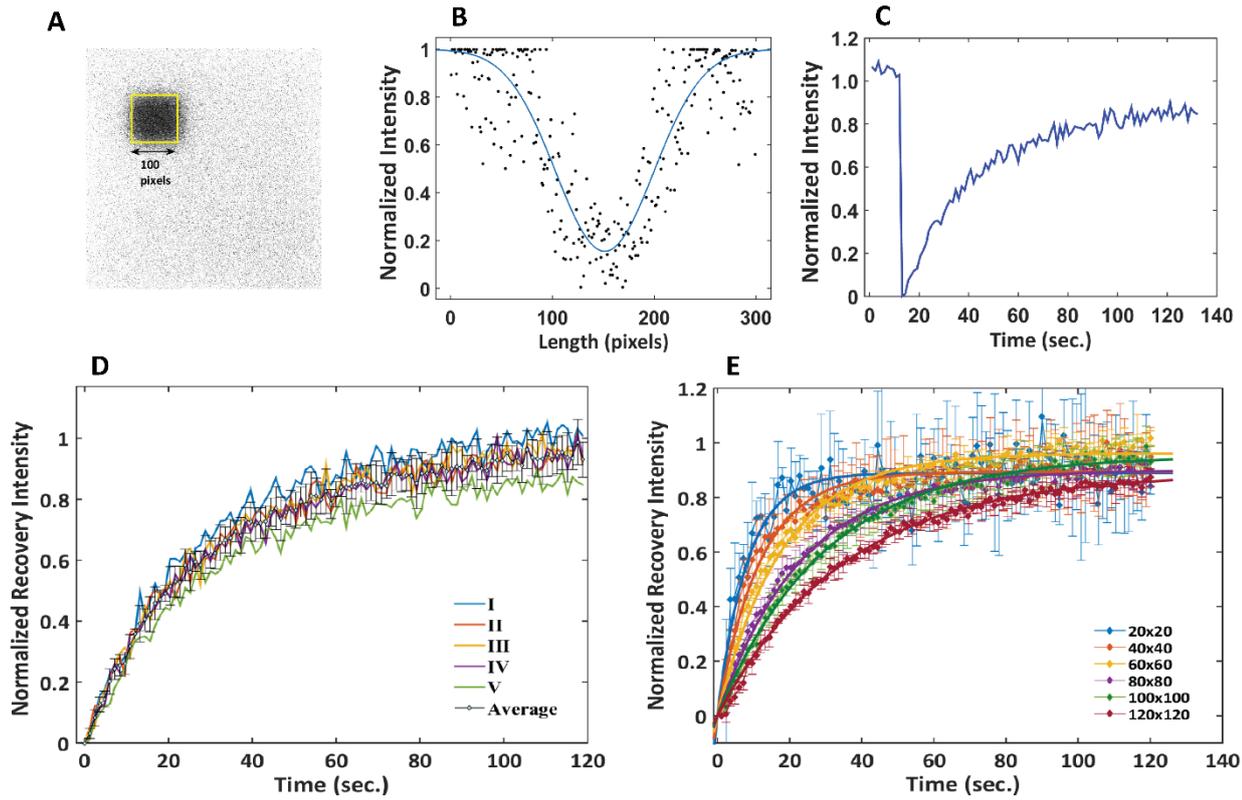

Figure 1 : Fluorescence Recovery After Photobleaching (FRAP). A. Confocal microscope fluorescence image for PEG solution with fluorescein, with the bleached spot shown as dark region, spanning about 100 pixels. B. The normalized fluorescence intensity profile of the bleached region can be fitted to a Gaussian with spot width w =96 ±5 pixels, corresponding to 39.7 ± 2 µm. C. A typical normalized fluorescence intensity profile obtained, with pre-bleach region (Intensity around 1), bleaching (Intensity = 0), and post bleach region, showing recovery. D. FRAP recovery curves obtained after bleaching different regions of 50 wt% PEG solution. E. FRAP recovery curves for 50 wt% PEG solution for different sizes of the bleached region given in the inset in pixels

In Equation 2, $I_N(t)$ is the time-dependent normalized post bleach intensity, $I_{data}$ is the mean intensity of the photobleached region, $I_0$ is the mean intensity of the photobleached region immediately after bleaching and $I_{whole}$ is the mean intensity of entire frame. A typical normalized intensity profile for 50 wt.% PEG (50P) is shown in Figure 1C. Unless heterogeneities were studied, the mean of five normalized post bleach profiles were calculated with the standard deviation as the error. This mean recovery profile was then considered for further analysis. Figure 1D represents the normalized recovery curves for five set of experiments performed for 50P and the average recovery curve. Figure 1E represents the effect of the size of the bleach ROI, 20x20, 40x40, 60x60, 80x80, 100x100 and 120x120 pixels ROI for 50P on their respective recovery profiles. Since some of the samples had bentonite with about 10 -20 µm clay particles, in order to get a comparative picture of the diffusion of fluorescein in the different PEG solutions, 100X100 and 120X120 pixels bleach regions are the better choices. For all the experiments, the size of the bleach region was chosen subsequently as 100x100 pixels.

Table 1 : Length scales (blob size $\xi$) and interaction strength ($\emptyset_{Dep}$, depletion interactions in presence of bentonite) for different concentrations of PEG aqueous solutions. The theoretical value of $R_g$ calculated from $R_g = 0.02 M_w^{0.58}$ = 6.246 nm and its experimental value is 4.3 nm[41]. The critical overlap concentration $c^* = \frac{M_w}{\frac{4}{3}\pi R_g^3 N_A}$ = 0.0289 g/ml.

| Sample | Volume Fraction (c in g/ml) | $\xi = R_g \left(\frac{c}{c*}\right)^{-0.76}$ (Theoretical in nm) | $\xi = R_g \left(\frac{c}{c*}\right)^{-0.76}$ (Experimental in nm) | $\xi = c^{\left(\frac{-\nu}{3\nu-1}\right)}$ (in nm) | $\frac{R}{\xi}$ (hydrodynamic radius of fluorescein, R=0.504 nm, $\xi$ from Col 4) | $\frac{R}{R_g}$ (Fluorescein) | $\frac{b}{R_g}$ (size of bentonite b = 25 µm) | $\frac{\emptyset_{Dep}}{k_B T} = -\frac{3b}{2R_g}c = -8.7 \times 10^3 c$ |
|---|---|---|---|---|---|---|---|---|
| 5P | 0.046 | 4.38721 | 3.020333 | 12.76143 | 0.165 | 0.116 | 5.8 X 10³ | 0.4 |
| 10P | 0.0986 | 2.457751 | 1.692016 | 6.793034 | 0.298 | | | 0.857 |
| 20P | 0.222 | 1.326344 | 0.913109 | 3.471891 | 0.55 | | | 1.93 |
| 30P | 0.38 | 0.881554 | 0.606897 | 2.225968 | 0.831 | | | 3.3 |
| 50P | 0.888 | 0.462477 | 0.318388 | 1.103221 | 1.58 | | | 7.65 |

In order to calculate diffusion coefficients of the solutions investigated, three different methods were employed and their results were compared. In the first method, the time dependent mean normalized recovery profile ($I_N(t)$) after photobleaching was fitted to Equation 3

$$I_N(t) = I_\infty(1 - e^{-(t/\tau)}) \quad (3)$$

where $\tau$ is the diffusion time, and $I_\infty$ is the intensity at which the FRAP recovery saturates. The results of the fit to Equation 3 for different PEG solutions are given in Table 2. From $I_N(t)$, the half recovery time ($t_{1/2}$) was calculated as the time at which $I_N(t)$ is $0.5 I_\infty$. In principle, the recovery curve does not reach the pre-bleach value in most cases due to immobile constituents in the solution, and the immobile fraction is the difference of the recovery curve from the pre-bleach value. The $t_{1/2}$ was then used to calculate the diffusion coefficient using Equation 4 and Equation 5[38,42]:

$$D = \frac{0.224 r_n^2}{t_{1/2}} \quad (4)$$

$$D = \frac{r_n^2 + r_e^2}{8 t_{1/2}} \quad (5)$$

In the above equations, $r_n$ is the actual radius of the photobleached region (50 pixels, or 20.7 μm) and $r_e$ is the effective radius of the photobleached region (91.35 pixels or 37.82 μm), which was calculated by fitting the postbleach recovery profile to a Gaussian (Equation 1). The D values calculated from Equation 4 (D1) and Equation 5 (D3) are given in Table 2.

D values were also obtained by fitting the normalized postbleach recovery profiles to Equation 6 (D2, Table 2)[43].

$$I_N(t) = I_{(\infty)}\left(1 - \sqrt{w^2(w^2 + 4\pi D t)^{-1}}\right) \quad (6)$$

where $I_N(t)$ is the mean normalized time dependent post bleach recovery fluorescence intensity, $I_{(\infty)}$ is the asymptotic fluorescence intensity, w is the half width of the bleached ROI (20.7 μm) and D is the diffusion coefficient. The fit results for FRAP data of all samples are given in Supplementary Table S1 and S2. As a control, FRAP data was collected from aqueous solution of fluorescein, the D values obtained from Equations 4, 5 and 6, and compared to the Stokes Einstein D value of fluorescein in water (with ~0.5 nm as the hydrodynamic radius of fluorescein in water[44]), which is calculated as 386.2 μm$^2$sec$^{-1}$ in the present experimental conditions. The D value obtained from Equation 5 matches this theoretical D value best (Table 2).

### 2.4. Phase contrast microscopy

Phase contrast images were captured with Nikon Eclipse TS100 microscope equipped with Nikon 40x-0.60NA phase contrast ELWD objective with 300 msec. exposure time. The camera used for imaging was Nikon DS Fi2 with resolution 1μm/pixel.

### 2.5. Structure factor and 2D correlation calculations

Microscopy gives an idea of structure and dynamics in real space, whereas scattering gives an idea of structuring in Fourier space, with respect to the scattering wave vector $q$. Both these techniques are complementary, but instrumentation required for simultaneous microscopy and scattering experiments is not trivial. This has led in recent years to the development of Differential Dynamic Microscopy (DDM)[45,46]. In this method, time series microscopy images are analyzed in Fourier space, using the principles of Fourier optics, removing the requirement of any sophisticated instrumentation. Although it has been used to probe structural dynamics, it results in a very sensitive detection of the static structure factor. If the intensity at a point **x'** in a microscope image in real space with coordinates (x', y') is represented by *I(x')*, for the image of size $N_x$ x $N_y$ pixels (=512 X 512 pixels for images in the present case) and magnification *M*, with $l_p$ (= 0.414 μm/pixel in our experiments) the size of each pixel, then the structure factor, which is the Fourier space ($q$ space, $q$ given by the Equation 7) representation of *I(x')* is given as $\tilde{I}(q)$ (Equation 8) [45]

$$q = \frac{2\pi M}{N_x N_y l_p} \quad (7)$$

$$\tilde{I}(q) = \sum_p I(x'_p) e^{-i(q \cdot x'_p)} \quad (8)$$

In Equation 8, the summation is over all the pixels, *p*, and **x'** or $(x'_p, y'_p)$ is the coordinate of the p$^{th}$ pixel. The above equation was used to calculate the structure factor from the confocal microscopy image of a sample (using the same setup as in FRAP experiments). Since imaging is done with a scanning confocal microscope, each image represents a thin section of the sample, so multiple scattering is negligible and the scattered fluorescence emission of the fluorescein is basically the scattered light captured, the transmitted illuminating beam is eliminated. The final static structure factor *S(q)* was obtained by averaging the structure factor (Equation 8) obtained from 10 frames for each sample. To calculate the spatial correlations in two dimensions, a convolution was performed over the product of image intensity matrix in Fourier space and its complex conjugate.

Table 2 : Results of FRAP data analysis for diffusion time (τ), diffusion half time (t$_{1/2}$), immobile fraction and diffusion constants (D1,D2, D3) calculated by three different methods as explained in the Materials and Methods section.

| Sample | PEG (wt%) | Bentonite (wt%) | CMC (wt%) | τ (sec) | t$_{1/2}$ (sec) | Immobile Fraction (M$_{If}$) | $D = \dfrac{0.224\, r_n^2}{t_{1/2}}$ (μm$^2$sec$^{-1}$) (D1) | D from Eq (6) (μm$^2$sec$^{-1}$) (D2) | $D = \dfrac{r_n^2 + r_e^2}{8 t_{1/2}}$ (μm$^2$sec$^{-1}$) (D3) |
|---|---|---|---|---|---|---|---|---|---|
| Water | 0 | 0 | 0 | 1.02 ± 0.04 | 0.70 ± 0.03 | 0 | 135 ± 6 | 401 ± 61 | 327 ± 15 |
| 5P | 5 | 0 | 0 | 0.99 ± 0.04 | 0.68 ± 0.03 | 0 | 139 ± 6 | 544 ± 97 | 337 ± 16 |
| 10P | 10 | 0 | 0 | 1.24 ± 0.07 | 0.85 ± 0.05 | 0.018 ± 0.005 | 111 ± 6 | 241 ± 29 | 270 ± 16 |
| 20P | 20 | 0 | 0 | 1.736 ± 0.009 | 1.20 ± 0.06 | 0.01 ± 0.003 | 79 ± 4 | 143 ± 22 | 193 ± 10 |
| 50P | 50 | 0 | 0 | 27.8 ± 0.7 | 19.2 ± 0.5 | 0.055 ± 0.007 | 4.9 ± 0.1 | 2.3 ± 0.1 | 12 ± 0.3 |
| 20P + 0.01B | 20 | 0.01 | 0 | 12.2 ± 0.2 | 8.4 ± 0.1 | 0.166 ± 0.003 | 25.5 ± 0.4 | 16.6 ± 0.2 | 87 ± 1 |
| 20P + 0.1B | 20 | 0.1 | 0 | 11.1 ± 0.1 | 7.7 ± 0.1 | 0.135 ± 0.002 | 12.3 ± 0.1 | 10 ± 1 | 29.9 ± 0.4 |
| 20 + 1B | 20 | 1 | 0 | 5.7 ± 0.2 | 4 ± 0.15 | 0.067 ± 0.003 | 23.9 ± 0.8 | 33 ± 4 | 57 ± 2 |
| 0.01B | 0 | 0.01 | 0 | 3.8 ± 0.1 | 2.6 ± 0.1 | 0.234 ± 0.004 | 35 ± 1 | 39 ± 6 | 87 ± 4 |
| 0.1B | 0 | 0.1 | 0 | 3.3 ± 0.1 | 2.3 ± 0.1 | 0.25 ± 0.004 | 40 ± 2 | 49 ± 7 | 98 ± 5 |
| 1B | 0 | 1 | 0 | 4.44 ± 0.01 | 3.08 ± 0.07 | 0.105 ± 0.003 | 70 ± 1 | 82 ± 1 | 241 ± 5 |
| 20P + 0.01C | 20 | 0 | 0.01 | 1.7 ± 0.04 | 1.18 ± 0.03 | 0.03 ± 0.004 | 81 ± 2 | 111 ± 13 | 196 ± 5 |
| 20P + 0.1C | 20 | 0 | 0.1 | 2.06 ± 0.05 | 1.42 ± 0.04 | 0.035 ± 0.004 | 67 ± 1 | 74 ± 9 | 162 ± 4 |
| 20P + 1C | 20 | 0 | 1 | 2.06 ± 0.06 | 1.42 ± 0.04 | 0.049 ± 0.004 | 67 ± 2 | 81 ± 9 | 162 ± 5 |
| 0.01C | 0 | 0 | 0.01 | 2.8 ± 0.1 | 1.9 ± 0.1 | 0.197 ± 0.003 | 48 ± 2 | 88 ± 10 | 117 ± 6 |
| 0.1C | 0 | 0 | 0.1 | 2.4 ± 0.1 | 1.7 ± 0.08 | 0.187 ± 0.005 | 56 ± 2 | 77 ± 7 | 135 ± 6 |
| 1C | 0 | 0 | 1 | 3 ± 0.1 | 2.1 ± 0.1 | 0.237 ± 0.03 | 44 ± 2 | 71 ± 12 | 108 ± 5 |

## 3. Results and discussion

### 3.1. Dynamics of fluorescent probes in the polymer matrix

A study of diffusion of fluorescent probes in solution using FRAP can be a sensitive measure of micro-scale mechanical properties as well as inhomogeneities, if any. This diffusion depends on the size of fluorescent probe particles (R) and the correlation length ξ of the polymer matrix which characterizes the size of the network formed by overlapping polymer chains in the semi-dilute regime[47]. If particle size is large, so that $\frac{R}{\xi} \gg 1$, then the particle diffusion is a probe of the macroscopic viscosity, as measured by a rheometer. On the other extreme, if $\frac{R}{\xi} \ll 1$, then the particle diffusion is mostly governed by the solvent viscosity, as the probes can easily navigate through the empty spaces in the polymer network. The intermediate regime of $\frac{R}{\xi} \sim 1$, the particles probe the local viscosity of the polymer matrix, and can be a sensitive tool to study micro-scale mechanical properties and heterogeneities.

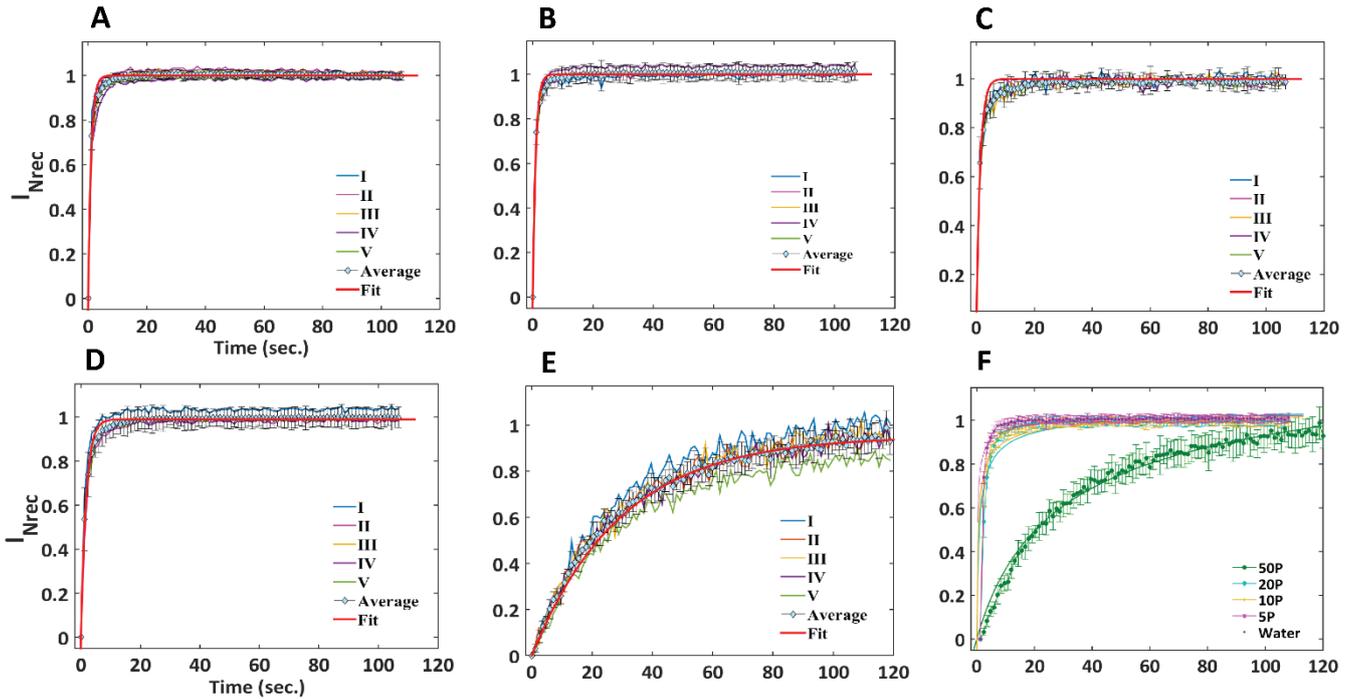

Figure 2 : FRAP recovery curves of normalized fluorescence intensity $I_{Nrec}$ as a function of time, for A. water; B. 5P solution; C. 10 P solution; D. 20 P solution; and E. 50 P solution, after bleaching five different regions in the sample, with each region represented with Roman letters from I to V. From these curves, the average curve is calculated and error bars are calculated as the deviation of the curves from the average value. The red curve is the fit with the Equation 3. The error bars increase with increase in PEG concentration, and for 20 P and 50 P solutions, the FRAP curves obtained in different regions have more deviations from the mean curve, with more immobile fraction (Table 2). F. Fits of average recovery curves to Equation 6, for solutions of different PEG concentrations as mentioned in the inset.

The intermediate regime was therefore chosen to probe micro-scale diffusion of fluorescein probes (hydrodynamic radius ∼ 0.5 nm), in 5P, 10P, 20P and 50P PEG solution, which have ξ varying between 3 to 0.3 nm (Table 1). As a control, FRAP experiment performed for fluorescein diffusion in water yielded a fast fluorescein diffusion time and diffusion constant very close to the value predicted by the Stokes Einstein relation (Figure 2A). Fluorescein diffuses fast in 5P solution also, with hardly any immobile fraction seen in the FRAP recovery curves (Figure 2B). Both the 5P and 10P solutions have similar recovery curves studied in different regions of the sample, so error in the average recovery curve is negligible (Figure 2B-C). However, the 20P solution had comparatively more scatter in the recovery curves obtained for different regions of the sample, and hence larger error in the average recovery, indicating the presence of aggregates (Figure 2D). The maximum scatter was observed in the 50P recovery (Figure 2E), which was also slower, demonstrating the enhancement of micro-scale inhomogeneities as the polymer concentration is spanned through the semi-dilute to the concentrated regime. The recovery curves are fitted well with an exponential function which gives the diffusion time (Figure 2F, Supplementary Table S1), and diffusion constants (D) are obtained both from this diffusion time (Equations 4 and 5) as well as directly from the fit to Equation 6 (Supplementary Table S2). The D values obtained from the FRAP recovery curves are given in Table 2. For the control sample, Equations 5 and 6 resulted in values of D closer to the theoretical Stokes Einstein value compared to Equation 4. In all cases, exponential function (Equation 2, Figure 2A-E) fitted the data better than the Equation 6 (Figure 2F), and therefore the $t_{1/2}$ obtained from the diffusion time τ was used for comparison between different solutions.

When bentonite particles are added to the 20P PEG solution, there is a huge (about 30%) scatter obtained in the recovery curves in different regions of the sample probed (Figures 3A-B), leading to variable immobile fractions much larger than those observed in the 20P solution, and large error bars in the average recovery profile (Figure 3A-B). The D values obtained from the mean recovery data from five different regions of the sample are given in Table 2, whereas the D values of the sampled regions separately and their respective immobile fractions are given in Supplementary Table S3. The heterogeneity in D values from different scanned areas of the PEG +B solutions in the microchannel is remarkable. Surprisingly, the mean D values for 20P+1B solution was found to be larger than the D values of 20P +0.1 and 20P+ 0.01B solution, and diffusion time was smaller indicating comparatively fast diffusion at the highest bentonite concentration studied. The heterogeneities point to enhanced propensity of bentonite aggregation at higher bentonite concentrations, possibly due to depletion interactions (analyzed in a subsequent section, with phase contrast microscopy images). Control experiments with only bentonite in solution showed faster diffusion (small τ) compared to the PEG + bentonite systems (Table 2, Supplementary Figure S1, Table S3).

It has been observed that aqueous solutions of NaCMC can have colloidally dispersed polyelectrolytes, or have networks and aggregates, depending on the degree of substitution[48,49]. SANS study of Lopez et al have found that NaCMC with DS=1.2 is molecularly dissolved in water with a locally stiff conformation in the semi dilute unentangled and entangled regimes, as well as the concentrated regime; they also found a tendency of weak aggregate formation due to the low-$q$ upturn of the scattering profiles[48]. In the present FRAP experiments, blends of NaCMC (DS=0.7) of varying concentrations (0.01C, 0.1C, and 1.0C) with the 20P solution showed diffusion time similar to that observed in the 20P solution with very less scatter in the average profile, but a larger amount of immobile fraction compared to that observed in 20P was found in the 20P+CMC systems (Figures 3C-D, Table 1). This can arise either due to the presence of weak aggregates, weak PEG-CMC structural complexes, or due to hindrance to the fluorescein diffusion due to the semiflexible nature of CMC chains. When control experiments were done with aqueous solution of only CMC, this immobile fraction was found to be greater than those observed in PEG+CMC solutions (Supplementary Figure S1, Table 2), pointing to two possible reasons- presence of CMC aggregates or the topological hindrance to fluorescein diffusion by semiflexible CMC. The D values for aqueous CMC solutions in the present case seem to be similar to the fast mode D-values obtained by dynamic light scattering experiments of CMC aqueous solutions, which also report additional slow and ultra-slow modes of diffusion[49]. The fast mode of diffusion in Ref 49 was independent of concentration, as we see in the present case, whereas the slow mode of diffusion was found to be strongly concentration dependent.

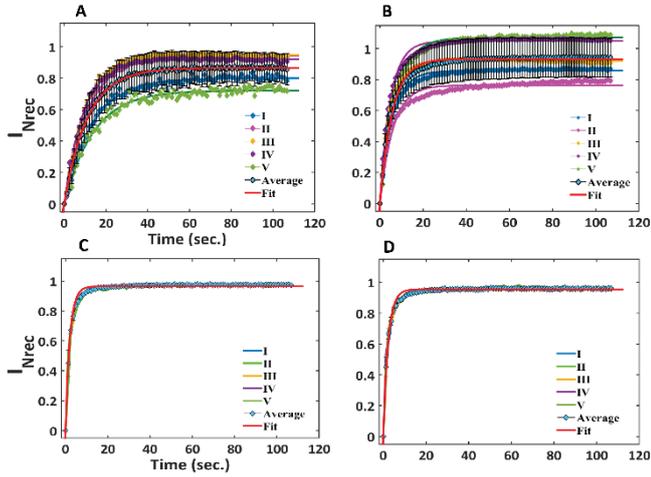
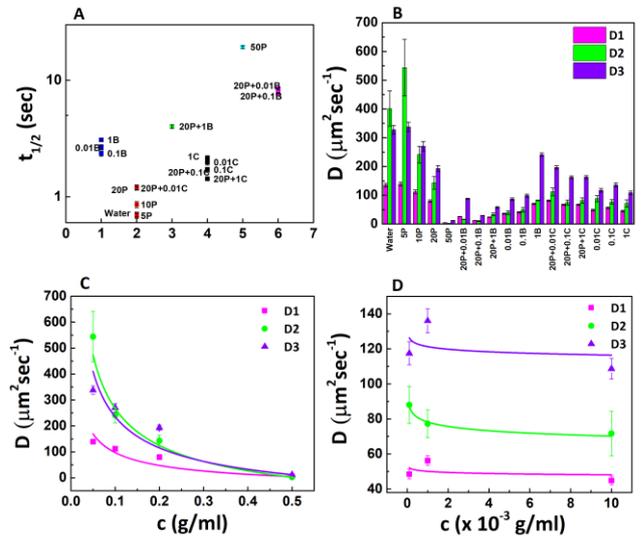

Figure 3 : FRAP recovery curves of normalized fluorescence intensity $I_{Nrec}$ as a function of time, for A. 0.1 wt% bentonite in 20 wt% PEG solution; B. 1 wt% bentonite in 20 wt% PEG solution; C. 0.1 wt% CMC in 20 wt% PEG solution; D. 1 wt% CMC in 20 wt% PEG solution, after bleaching five different regions in the sample, with each region represented with Roman letters from I to V. From the data, the average (blue diamond) is calculated and error bars are calculated from the deviation of the curves from the average value. All curves are fits of data points with the Equation 2. Bentonite solutions have a high value of immobile fraction with large spread in the FRAP recovery curves, which get enhanced with increase in bentonite concentration. Error bars of the average FRAP recovery are therefore large for these bentonite systems. In case of CMC solutions, there are immobile fractions, but the spread in FRAP curves obtained by photobleaching different regions is almost negligible, indicating high level of homogeneity in CMC solutions compared to bentonite solutions.

Figure 4 : A. Cluster graph of diffusion half time of fluorescein molecule in different solutions. Each cluster is shown with a specific colour. The diffusion time slows down more in the presence of bentonite than in presence of CMC in PEG matrix. B. Diffusion coefficients calculated from the three different approaches, as given in the text and in the Table 2. C. Diffusion coefficient of fluorescein in PEG matrix for different concentrations of PEG, fitted to the non-entangled Rouse scaling $D \sim c^{-0.54}$. D. Diffusion coefficients as a function of concentration for the pure CMC solution, which obey the pure polyelectrolyte scaling relation, $D \sim c^0$, with D independent of concentration.

Overall, a cluster analysis of the diffusion half time $t_{1/2}$ shows that 10P, 5P cluster together with water, with 20P and 20P+0.01C similar to this cluster (Figure 4A). However, 20P+ higher concentrations of CMC form separate cluster. This cluster has a lower $t_{1/2}$ compared to the cluster formed by 20P+0.01B and 20P+0.1B. 20P+1B forms a separate cluster with a lower $t_{1/2}$ compared to the other PEG-bentonite solutions, implying the presence of aggregates, and underlining the need for studying solution structure details. For comparison, the mean D values calculated from FRAP data for PEG, PEG+B and PEG+C solutions with three different equations (Equations 4 -6, Methods section) are shown as bar plots in Figure 4B. Since there are several length scales present in the semi-dilute and concentrated PEG, PEG+B and PEG+CMC, application of the Stokes Einstein equation to get an idea of the theoretical value of D can lead to improper estimates. For the PEG solutions, the mean D values follow Rouse scaling as a function of concentration (c, in volume fraction) $D \sim c^{-0.54}$ for c < 0.8 (Figure 4 C), whereas the solutions with only CMC have concentration independent scaling of mean D values, with $D \sim c^0$, the typical theoretical polyelectrolyte scaling (Figure 4D).

### 3.2 Response of PEG solutions under oscillatory shear as a function of strain amplitude

A strain sweep under oscillatory shear not only gives us an estimate of linear viscoelastic regime, but is also a sensitive test of presence of shear-induced microstructure formation and microstructure response under large amplitude oscillatory shear in complex fluids[50]. In case of PEG solution in water, the linear viscoelastic regime persists till the largest strain amplitude of 100% investigated, for all concentrations except 50P, which is the highest concentration examined. The 50P solution exhibits an enhancement in shear modulii (and the complex shear modulus $G^* = \sqrt{G'^2 + G''^2}$, Figure 5A) from a strain amplitude of about 10% onwards, possibly due to the presence of hydrogen bonded microstructures which resist deformation until about 50% strain, and then align in the direction of flow to lead to a decrease in the complex modulus for larger strain amplitudes. Such a microstructure-induced overshoot is not observed in the other less concentrated PEG solutions. The addition of bentonite to the PEG solution (PEG + B, Figure 5B) leads to an enhancement in both the storage and loss modulii, and this feature is bentonite concentration dependent (Supplementary Figure S2). However, the remarkable feature in PEG+B solutions is that the linear viscoelastic regime shortens with increase in bentonite concentration and a continuous decrease in shear modulus is observed for 1% bentonite (20P+1B, Figure 5B) with increase in strain amplitude, demonstrating the presence of microstructures which slide in the direction of shear in PEG+B solutions, with subsequent softening and yielding. Addition of CMC to PEG solution leads to an enhancement in the complex shear modulus (Figure 5C), but the response is largely linear until the highest strain amplitude of 100% investigated. At strain amplitudes below 0.1%, storage modulus is noisy, but a weak strain-induced thinning is observed for PEG+CMC, indicating breaks in weak PEG-CMC structural complexes (Supplementary Figure S2).

### 3.3. Dynamic scaling of shear modulii in the presence of interactions

At length scales smaller than the size of the polymer chain, relaxation is subdiffusive, i.e. either Rouse or Zimm, with time scales longer than the monomer relaxation time ($\tau_O$), but shorter than the Rouse or Zimm relaxation time. On time scales longer than the Rouse ($\tau_R$) or Zimm ($\tau_Z$) relaxation time, the chain relaxation is diffusive. For frequencies $\frac{1}{\tau_R} \leq \omega \leq \frac{1}{\tau_O}$, the relaxation dynamics follows the Rouse model, with the scaling relation $G'(\omega) \approx G''(\omega) \approx \omega^{1/2}$, if hydrodynamic, excluded volume, or other interactions are effectively screened[51]. In the presence of hydrodynamic interactions, the Zimm model holds good, with a scaling relation $G'(\omega) \approx G''(\omega) \approx \omega^{2/3}$. In the presence of both hydrodynamic or excluded volume interactions, the scaling relation $G'(\omega) \approx G''(\omega) \approx \omega^{5/9}$ is usually followed[51]. At low frequencies, with $\omega < \frac{1}{\tau_R}$, the typical viscoelastic response of a liquid with $G' \sim \omega^2$ and $G'' \sim \omega$ should be observed. The Rouse model specifically, has been formulated for single polymer chains, and effect of inter-chain interactions has been taken into account in the form of a constant friction representing intermolecular interactions[51]. Hence, if there are random forces acting on polymer segments, the Rouse model has been observed to fail to give a proper description of polymer dynamics, e.g. for polymer dynamics near surfaces, interfaces, in the presence of associative forces such as hydrogen bonding (H-bonds), or due to structural alpha relaxation which induce dynamical heterogeneities[9,52–55]. The presence of weak associations such as Van der Waals or H-bonds have interaction energies of the order of few $k_BT$, leading to reversible association-dissociation dynamics which influence relaxation and diffusion as a result of which apart from the Rouse time $\tau_R$, an additional time scale due to bond lifetime, $\tau_b$ is introduced. This has been found to have an effect on the $G'(\omega)$ scaling, with a power law exponent of $\omega$ higher than that predicted by the Rouse model[13]. Apart from the time scales, the different length scales present in the system also influence the dynamics.

A crucial parameter which affects the length scales in the system is the concentration at which chains touch each other, or the overlap concentration for a neutral polymer in a good solvent is given by the expression $c^* = \frac{3M}{4\pi N_A R_g^3}$ where M is the molecular weight (20000g/mol for PEG), $N_A$ is the Avogadro number. $R_g$ is the radius of gyration of PEG in water, and has been experimentally found to be 4.3 nm from SANS experiments on M=20,000 PEG dissolved in heavy water. This yields $c^* = 0.0289\ g/ml$ for M=20000 PEG. From this calculated value of $c^*$, we find that the 5P sample with $c/c^* \approx 1.1$, is just around the overlap concentration at the semi-dilute limit which includes 10P, 20P and 30P also. For 50P sample, $c/c^* \approx 3$, the chains overlap and could possibly entangle. The de Gennes correlation length ξ at which the interactions between chains are screened out, can be calculated using the expression $\xi \sim c^{\frac{-\nu}{3\nu-1}}$ (Table 1), where c is the concentration in g/ml and the exponent ν = 0.588 for a neutral polymer in a good solvent. The correlation length is maximum for 5P at ξ ≈ 12.7 nm, and minimum for 50P (ξ ≈ 1.1 nm), indicating that all electrostatic and excluded volume interactions are screened out as ξ is smaller than the size of a molecule for 50P. In case of 10P, 20P and 30P, ξ ≈ 6.8, 3.5, and 2.2 nm respectively (Table 1). Additional length scales due to associative clusters and depletion interactions if introduced in the system, could potentially influence the scaling of the storage modulus. Simulations of Prabhakar et al show that chain stretching weakens screening of hydrodynamic interactions in the semi-dilute limit in theta solvents, which can also influence the screening lengths in the system[56,57].

In order to investigate the effect of inhomogeneities and inter-molecular interactions on the scaling of the oscillatory storage modulii, we have studied the mechanical response of PEG solutions from three perspectives – (i) effect of polymer concentration-dependent microstructures due to dynamic H-bonds or short ranged interactions, (ii) effect of the presence of polyelectrolytes or rigid chains with screened Coulomb interactions, and (iii) effect of the presence of nanoparticles such as bentonite clay, introducing depletion interactions.

### 3.3.1. Effect of PEG solution microstructure on Rouse relaxation dynamics

In aqueous solutions, simulations reveal that PEG molecule adopts a helical semi-flexible structure with hydrophobic and hydrophilic patches on the surface, with surface hydrophobicity enhanced with increase in molecular weight and concentration[58,59]. Since there are both hydrophilic and hydrophobic domains in PEG, together with H-bonding interactions with the solvent and among the polymer chains themselves, concentration-dependent microstructure in aqueous solution is possible, as observed in various scattering studies. Response of PEG microstructures in solution to small amplitude oscillatory shear (SAOS) can reveal information about microstructural stability and the overall relaxation dynamics. We have investigated the SAOS response at 0.1% (Figure 6) and 1% strain (Supplementary Figure S3) for a series of concentrations of PEG in water, which are referred in the text following the notation given in Table 2 as 5P, 10P, 20P, 30P and 50P.

The storage modulii ($G'(\omega)$) for all the PEG solutions resemble the terminal flow of viscous fluids at low frequencies, and have a $\omega^2$ dependence for $\omega < \tau_d^{-1}$, with $\tau_d$ the terminal relaxation time. This is followed by short elastic plateau for $\tau_d^{-1} \leq \omega \leq \tau_e^{-1}$ (Figure 6A for 0.1% strain, Table 3, Supplementary Figure S3A for 1% strain, Supplementary Table S4). For the regime $\omega > \tau_e^{-1}$, the scaling is a concentration-dependent power law $\omega^\alpha$, with α greater than the Rouse scaling of 0.5 (Table 3). The exponent α is the highest for the 10P solution (Table 3). α has a non-trivial dependence on the concentration of PEG and the strain amplitude.

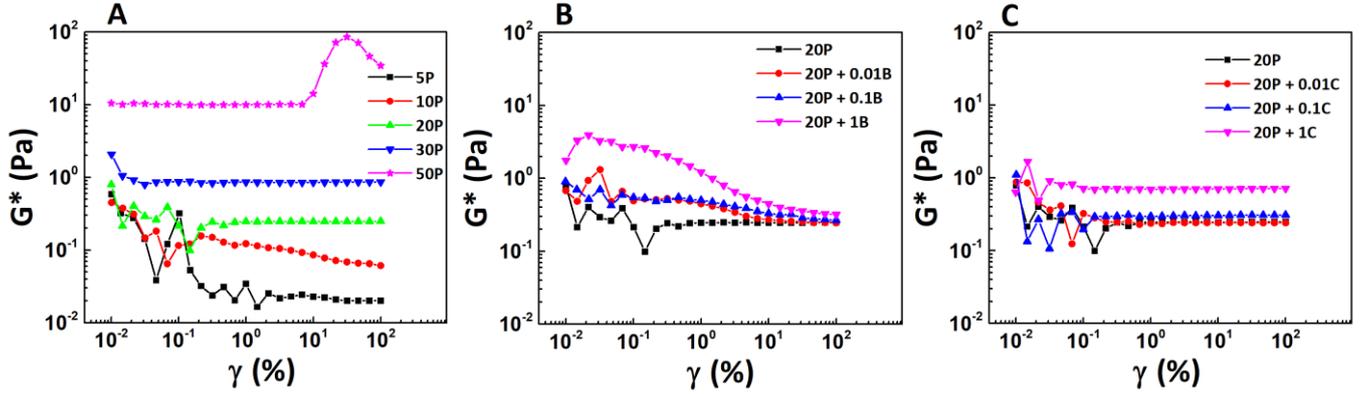

Figure 5: Response to oscillatory shear as a function of strain amplitude of A. different concentrations of PEG solutions from the lowest, 5P, to the highest, 50P. The 50P solution exhibits an overshoot due to stiffening, followed by yielding, illustrating the presence of microstructures. B. 20P mixed with different concentrations of bentonite (B). The highest concentration of B (1B) starts yielding from a strain amplitude of even less than 0.1%, again indicating structuring in the solution. C. 20P mixed with different concentrations of CMC. Apart from the noisy behaviour at very low amplitudes, the response is linear in the strain amplitude range investigated.

The 5P solution with PEG concentration just bordering the overlap concentration (semi-dilute) has α ~0.5, or the typical Rouse behaviour, implying that hydrodynamic and excluded volume interactions are screened. Increase of PEG concentrations increases associative interactions such as H-bonds which also have time-dependent association-dissociation kinetics. The presence of associative interactions could imply the presence of an additional time scale, leading to scaling exponent greater than the Rouse exponent in the range $\tau_e^{-1} \leq \omega \leq \tau_b^{-1}$, and this range is concentration dependent due to the inherent concentration dependence in the kinetics[10]. A closer examination of the scaling beyond the elastic plateau reveals this additional timescale $\tau_b$ (inset of Figure 6A), beyond which the Rouse relaxation dominates. This transition to the Rouse regime is not observed in the frequency range investigated in the present experiments for 10P. Since $\tau_R$ is also concentration dependent, it increases with increase in PEG concentration and commencement of Rouse behaviour shifts to lower frequencies, leading to a shortening of the range $\tau_b^{-1} \leq \omega \leq \tau_R^{-1}$, until $\tau_R \approx \tau_b$ in the highest concentration, 50P, for 1% strain amplitude (Supplementary Figure S3). This scaling behaviour of the storage modulii and the concentration dependence of $\tau_b$ is illustrated in the form of a schematic diagram (Supplementary Figure S4).

The loss modulii in both dilute and semi-dilute regime shows an upturn at low frequencies implying slow microstructural relaxations (Figure 6D for 0.1% strain and Supplementary Figure S3D for 1.0 % strain) for 5P, 10P and 20P. The mid-frequency region has another prominent peak in $G''$, indicating fast structural relaxations of around 0.1 sec., for concentrations 5P, 10P and 20P; but this peak reduces in 30P and disappears altogether in 50P, which has a predominantly lossy behaviour. In the case of 50P, structural relaxations would possibly have become slower and shifted to lower frequencies, not accessed in the present case. For the 20P, 30P, and 50P solutions, low frequency scaling of $G''\sim\omega$ is observed, similar to the liquid viscoelastic response. For the 5P and 10P solutions, the large microstructure relaxation peak present in low frequencies impedes the observation of the terminal viscoelastic fluid scaling. Moreover, the high frequency scaling of $G''(\omega)$ does not match that of $G'(\omega)$ (see Supplementary Table S4), a phenomenon which has been observed in many polymer solutions[60]. As long as the applied forces are smaller than the forces holding the microclusters and cluster networks together, elastic energy is stored, $G'$ is greater than $G''$, as seen for the dilute and semi dilute regime for 5P, 10P, 20P, and 30P. With increase in ω, there is cluster densification under shear leading to the onset of entanglements for $\tau_d^{-1} \leq \omega \leq \tau_e^{-1}$, producing a plateau in the G' (Figure 6A for 0.1% strain and Supplementary Figure S3A for 1.0 % strain). Even if there are no entanglements, the plateau could be due to the densification of networks formed by the overlapping chains.

### 3.3.2. Effect of the presence of CMC on the oscillatory shear response of PEG solution

The degree of substitution (DS) of carboxyl groups in NaCMC have been found to play an important role in CMC dispersion in aqueous solution. SANS studies on aqueous solutions of NaCMC with a DS=1.2 has found CMC to be molecularly dispersed with indications of presence of aggregates, whereas light scattering and rheology studies on aqueous solutions of NaCMC with DS=0.8 show the presence of colloidally dispersed CMC aggregates in aqueous solution[49,61]. In the latter case, the authors have reported the critical overlap concentration or the semi-dilute non-entangled concentration, c*, to be between 7.1 x$10^{-5}$ and 4.6 X $10^{-4}$ wt%, entanglement concentration $c_e$ between 0.11 and 0.16 wt % and concentrated regime c**, when the correlation length becomes smaller than the thermal blob size, to be between 0.46 and 0.45 wt%, with the help of viscosity measurements. In the present experiments, we have studied NaCMC concentrations of 0.01 wt%, 0.1 wt% and 1.0 wt% with NaCMC of DS≈0.7. Hence approximately, the concentrations 0.01 and 0.1 fall between the overlap concentration c* and the entanglement concentration $c_e$, whereas the 1.0 wt% solution lies in the concentrated regime, where all excluded volume interactions are screened and chains are expected to behave as ideal chains.

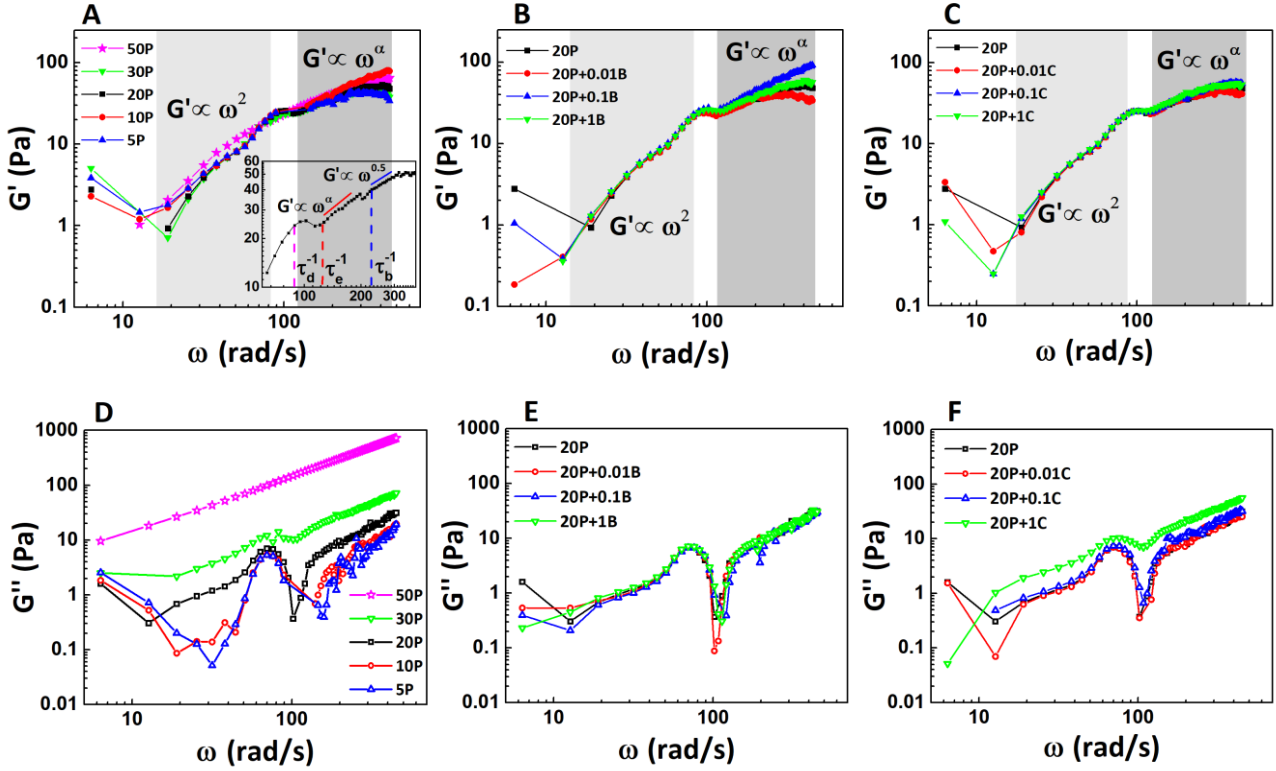

Figure 6 : Storage (G') and loss (G") modulii as a function of the frequency for 0.1% strain amplitude. Three regimes are observed in the storage modulii – a $\omega^2$ scaling regime at $\omega < \tau_d^{-1}$ showing viscoelastic fluid type terminal relaxation (light grey), an elastic plateau independent of $\omega$ for $\tau_d^{-1} \leq \omega \leq \tau_e^{-1}$, followed by a segmental relaxation regime with $\omega^\alpha$ scaling (dark grey). The $\alpha$ values are given in Table 3 for all samples investigated. A. G' for different concentrations of PEG solutions. A closer examination of the 20PEG scaling in the inset reveals the presence of an additional time scale $\tau_b$, and $\tau_d^{-1}$=89 rad/s, $\tau_e^{-1}$=121 rad/s, $\tau_b^{-1}$= 228 rad/s. For $\omega > \tau_b^{-1}$, the power law behaviour exhibited is the Rouse scaling of $\omega^{0.5}$. B. G' for 20P solution mixed with three different concentrations of bentonite (B), and C. G' for 20P solution mixed with three different concentrations of CMC. The corresponding G" are shown in panels D, E, and F. G" shows an upturn at low frequencies indicative of microstructural relaxations for PEG solutions less than 50P concentration (D), and an intermediate peak around 70 rad/s, which persists in the presence of bentonite (E) as well as CMC (F).

Oscillatory shear of the 1.0 wt% NaCMC aqueous solution show typical viscoelastic fluid behaviour with $G'\sim\omega^2$ and $G''\sim\omega$ scaling for frequencies below the inverse of the terminal relaxation time, $\tau_d^{-1}$. Following this, there is a short entanglement plateau between $\tau_d^{-1} \leq \omega \leq \tau_e^{-1}$ (Figure 6B, Table 3). For frequencies greater than $\tau_e^{-1}$, the ideal chain Rouse scaling of $G'\sim\omega^{0.5}$ is observed. Similar scaling behaviour of storage modulus is observed when 1 wt% NaCMC is mixed with 20P (20P+1C, Table 3, Figure 6B). When the NaCMC concentration is decreased in 20P, i.e. for 20P+0.01C and 20P+0.1C, the scaling of storage modulii is Rouse-like for frequencies greater than $\tau_e^{-1}$ at 0.1% strain amplitude. For higher strain amplitude of 1.0%, there is again a departure from the Rouse behaviour with scaling exponent of storage modulii similar to that observed in the 20P solution (Supplementary Figure S3B). Amplitude sweep experiments show that linear viscoelastic regime is present at 1% strain in all the 20P+CMC solutions. Hence this departure from the ideal Rouse behaviour could be due to dominant associative interactions in the 20P, despite of the presence of CMC of concentration greater than the overlap concentration. At 0.1% strain, the entanglement plateau modulus was found to be similar to the 20P, but a marked difference is that the loss modulus increases more rapidly with increase in CMC concentration, and a crossover is observed at 482 rad/s for 20P+1C, when $G' = G''$. For 1.0% strain, enhancement for $G''$ for 20P+1C results in a crossover which is earlier, at 451 rad/s (Supplementary Figure S3). This crossover frequency implies a relaxation time of around 0.002 sec., possibly of weak PEG-CMC associative networks

.3.3.3 Effect of the presence of Bentonite nanoparticles on the oscillatory shear response of PEG solution

Incorporation of clay nanoparticles in polymer matrix has been known to enhance mechanical, optical or thermal properties, due to microstructural changes as a result of polymer-clay interactions[19,62,63]. In aqueous solutions, any clay or bentonite nanoparticles have a layered silicate structure with metal ions chelated on the inner surface. PEG, as well as polyethylene oxide, is known to get adsorbed on the surface of these clay particles[63]. This results in supramolecular organizations due to bridging of neighbouring clay particles by polymer chains. Rheo SANS experiments on bentonite polymer aqueous suspensions have found that introduction of oscillatory shear first results in orientation of bentonite particles, followed by elastic stretching of surface adsorbed polymers leading to enhancement in $G'$ with shear[31]. For a fixed polymer concentration (20P), the effect of addition of small amount of bentonite (0.01 wt% (0.01B), 0.1 wt%(0.1B) and 1.0 wt%(1B)) was first studied by strain sweep experiments (Figure 5B). Addition of 0.01% and 0.1% bentonite slightly improved the complex modulus G*, but resulted in an onset of non-linearity from about 1% strain amplitude in comparison to the pure 20P solution, which has a linear response in the amplitude range investigated (Figure 5B).

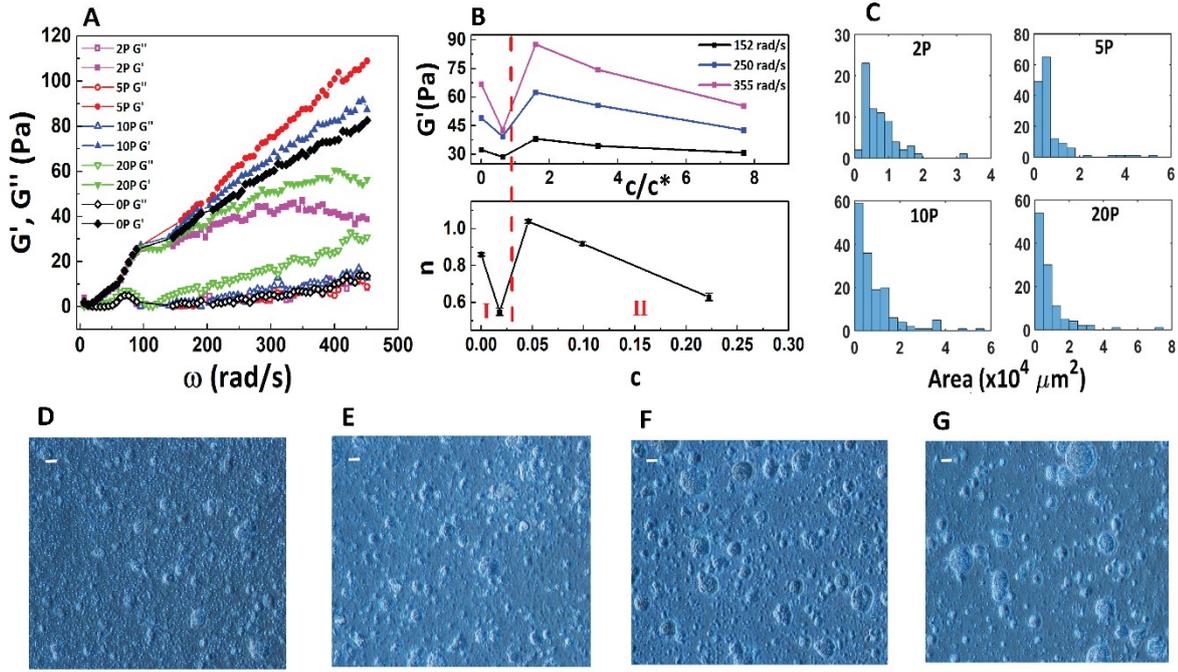

Figure 7 : A. Storage and loss modulii at 0.1% strain amplitude for different concentrations of PEG in 1wt% bentonite (1B), as shown in the inset, with 0P indicating the solution of only bentonite and water. B. Concentration dependence of G', with the concentrations in volume fraction (c) scaled with respect to the critical overlap concentration, $c^*$. The bottom panel shows the concentration dependence of the power law exponent n for $\omega^n$ scaling of G' shown in A, at frequencies higher than the elastic plateau region. Clearly, there is a departure from the Rouse exponent of n=0.5. C. Distribution of the area of the bentonite aggregates for different concentrations of PEG, calculated from the corresponding phase contrast images shown in panels D, E, F and G. Increase in PEG concentrations leads to enhancement of aggregate size.

Addition of 1% bentonite (20P+1B) resulted in a five-fold enhancement in G*, but the onset of non-linear response and yielding started much earlier, from about 0.5% strain amplitude onwards (Figure 5B). Frequency sweep experiments performed for these four samples at 0.1% strain result in a terminal flow regime, with $G'\sim\omega^2$ and $G''(\omega)\sim\omega$, with the storage modulus G' greater than G", implying that the characteristic relaxation time ($2\pi/\omega$) has occurred at a frequency lower than the investigated frequency window (Figure 6B). This is followed by a short elastic plateau. For the 20P+0.01B, a second crossover seems to occur around 450 rad/sec, not observed in the 20P solution, indicating a possible relaxation of surface adsorbed PEG on bentonite layers. Addition of bentonite does not seem to alter G"(Figure 6E), but it does alter the variation of G' with ω beyond the elastic plateau.

Table 3 : The values of power law scaling exponent α in $G' \propto \omega^\alpha$ for 0.1% and 1.0% strain amplitude in two different frequency regimes shown as shaded region in Figure 6 for 0.1% strain and Supplementary Figure S3 for 1% strain, with α values close to the Rouse model value of 0.5 highlighted in bold.

| Sample | Volume Fraction of PEG (c) | α for $\omega < \tau_d^{-1}$ (light grey, Figure 6) 0.1% | α for $\omega > \tau_e^{-1}$ (dark grey, Figure 6) 0.1% | α for $\omega < \tau_d^{-1}$ (light grey, Suppl. Figure S3) 1.0% | α for $\omega > \tau_e^{-1}$ (dark grey, Suppl. Figure S3) 1.0% |
|---|---|---|---|---|---|
| 5P | 0.046 | 1.64 ± 0.05 | **0.52 ± 0.03** | 2.07 ± 0.11 | **0.55 ± 0.01** |
| 10P | 0.0986 | 1.71 ± 0.05 | 0.87 ± 0.01 | 2.04 ± 0.09 | 0.95 ± 0.007 |
| 20P | 0.222 | 1.99 ± 0.08 | 0.56 ± 0.02 | 2.05 ± 0.1 | 0.72 ± 0.01 |
| 30P | 0.38 | 2.11 ± 0.11 | **0.52 ± 0.03** | 2.01 ± 0.08 | **0.51 ± 0.02** |
| 50P | 0.888 | 1.92 ± 0.02 | 0.63 ± 0.01 | 1.52 ± 0.02 | **0.52 ± 0.01** |
| 20P+0.01C | 0.222 | 2.05 ± 0.09 | **0.52 ± 0.02** | 2.03 ± 0.09 | 0.85 ± 0.009 |
| 20P+0.1C | 0.222 | 1.88 ± 0.05 | **0.52 ± 0.02** | 1.98 ± 0.09 | 0.68 ± 0.01 |
| 20P+1C | 0.222 | 1.85 ± 0.04 | **0.54 ± 0.02** | 1.99 ± 0.08 | **0.54 ± 0.02** |
| 1C | 0 | 1.90 ± 0.05 | **0.53 ± 0.03** | 2.02 ± 0.09 | **0.52 ± 0.01** |
| 20P+0.01B | 0.222 | 1.86 ± 0.05 | 0.57 ± 0.04 | 2.06 ± 0.1 | 0.69 ± 0.01 |
| 20P+0.1B | 0.222 | 1.85 ± 0.04 | 0.92 ± 0.01 | 1.92 ± 0.07 | 0.68 ± 0.01 |
| 20P+1B | 0.222 | 1.83 ± 0.04 | 0.63 ± 0.02 | 2.03 ± 0.1 | 0.65 ± 0.01 |
| 1B | 0 | 1.70 ± 0.04 | 0.861 ± 0.009 | 2.07 ± 0.08 | 0.54 ± 0.01 |
| 2P + 1B | 0.022 | 2.11 ± 0.16 | 0.55 ± 0.02 | 2.04 ± 0.1 | 0.69 ± 0.01 |
| 5P+1B | 0.046 | 2.07 ± 0.08 | 1.04 ± 0.01 | 2.07 ± 0.1 | 1.02 ± 0.002 |

In case of 20P+0.01B, the G' dependence was similar to the 20P solution, with Rouse-like $G' \propto \omega^{1/2}$ for 0.1% strain(Figure 6A). Increase in the amount of bentonite (0.1B) leads to higher values of G' from ω=100 rad/s onwards, possible due to formation of more polymer-clay supramolecular structures, which orient at higher shear rates in addition to stretching of polymer chains adsorbed on the clay surface, as observed in SANS studies of PEO-laponite. G' shows a near-linear ($G' \propto \omega^{0.9}$) dependence on ω for 20P+0.1B. Enhancement of bentonite concentration to 1% (1B) results in decrease of G' values at frequencies higher than 150 rad/s and $G' \propto \omega^{0.6}$ behaviour (Figure 6A, Table 3). Experiments on laponite particles in PEO also show a decrease in the G' with more addition of laponite particles in a concentrated polymer matrix due to the increase in the density of polymer-clay bridging interactions and an enhancement in the relaxation time due to ageing effects and gelification[23,27].

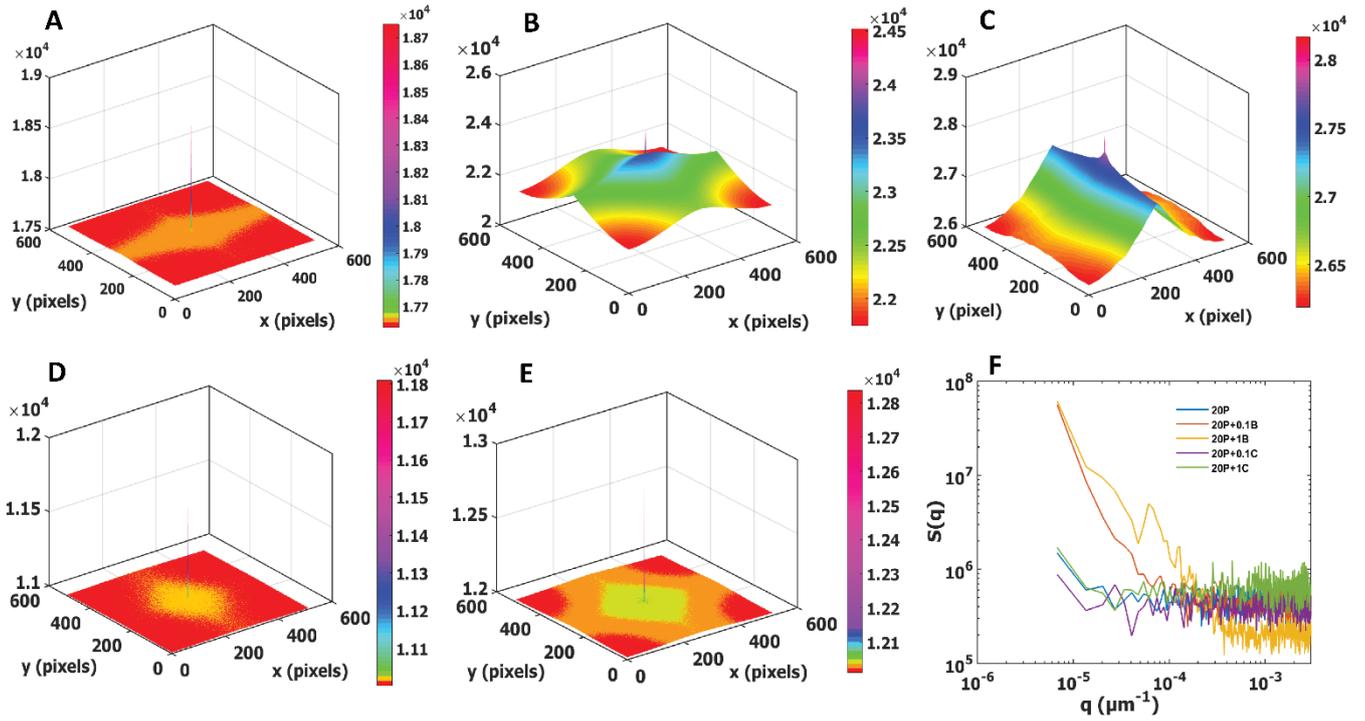

Figure 8 : 2D spatial correlation functions for A. 20 wt% PEG solution; B. 20 wt% PEG + 0.1 wt% bentonite; C. 20 wt% PEG + 1 wt% bentonite; D. 20 wt% PEG + 0.1 wt% CMC solution, E. 20 wt% PEG + 1 wt% CMC solution, indicating longer ranged correlations in solutions containing bentonite, possibly due to bentonite aggregates. F. Static structure factor $S(q)$ as a function of wave vector $q$. In the case of solutions with bentonite, there is an enhancement of $S(q)$ in the low $q$ region indicating inhomogeneities or aggregates.

In order to investigate further this decrease in G' at 1% bentonite, we decided to perform a series of experiments keeping the bentonite concentration fixed at 1% and gradually increasing the polymer concentration from 2% - 20% (2P -20P). We find that addition of a small amount of PEG (2% PEG, 2P) to the bentonite solution does decrease the G', but further addition of PEG in case of 5P and 10P, there is an enhancement in G' (Figure 7A-B). However, for a fixed frequency, beyond the critical overlap concentration of the polymer (c/c*=1), there is again a decrease in the G' values (Figure 7B). Phase contrast images of these samples at zero shear conditions show the evidence of enhanced flocculation of bentonite particles, with increase in polymer concentration (Figure 7C-G).

With increase in polymer concentration, there is more non-adsorbed polymer chains, and this can result in entropic attractive depletion forces between colloidal particles. The size of the depletion zone is approximately equal to the radius of gyration of the polymer, and if two colloidal particles approach closer than twice the size of the depletion zone, then there is no polymer in the depletion zone (configurational distortion of polymer in this small region leads to high entropic cost), and a net attractive force brings the colloidal particles together. The strength of the depletion force depends therefore on the concentration of polymer solution, i.e. the osmotic pressure of the solution, and the range of the force depends on the size of the polymer molecule. Tuning this interaction has been found to yield interesting phase separation kinetics or the presence of long-lived metastable glassy and gel states in colloid-polymer mixtures[25,36]. Possibly this could result in phase separated polymer-rich and bentonite-rich states at high shear rates leading to decrease in G' values in high polymer concentrations. An approximate estimation of the depletion interaction potential is given in units of $k_B T$ by $\frac{\emptyset_{Dep}}{k_B T} = -\frac{3b}{2R_g}$, where $b$ is the radius of the colloidal particle (≈25 μm for bentonite particles measured by phase contrast microscope). The calculated PEG-bentonite depletion potentials are large (Table 1), implying that depletion interactions can lead to polymer-rich and bentonite-rich phases as PEG concentration is increased. For 20P, the depletion interaction which tends to cause phase separation can compete with adsorption of PEG on clay surface which tends to form PEG-clay associated structures, leading to breakage of polymer-clay bridged networks. This can lead to heterogeneities in the system and non-linearities in oscillatory shear modulii with increase in strain amplitude, which gets aggravated with increase in bentonite concentration, a fact which is reflected in the FRAP and amplitude sweep experiments.

### 3.4. Static structure of PEG, PEG + Bentonite and PEG + Cellulose

To study structural information arising from interactions in the systems, two-dimensional correlation functions and the corresponding structure factors in the Fourier space were calculated for the PEG, PEG + cellulose and PEG + bentonite systems. Several groups have shown that by analysing the intensity of microscope images in Fourier space, one can extract static as well as dynamic structure factors same as the ones obtained in dynamic light scattering experiments. Microscopy methods have been found to give structural information for a wider range of *q*-values or length scales compared to dynamic light scattering. Moreover, highly concentrated solutions which become difficult to study by dynamic light scattering due to multiple scattering can be easily studied by microscope image analysis.

According to the convolution theorem, the Fourier transform of the correlation function or the convolution is the product of the Fourier transform of the image with the Fourier transform of its complex conjugate. Hence, the 2D autocorrelation function is obtained by an inverse Fourier transform of the convolution of the 2D images obtained by confocal microscope using fluorescent molecular probes in the different solutions investigated in the previous sections. The results are shown in Figure 8A-E. Clearly, from this figure, the highest degree of correlations are found in the PEG-bentonite system, as a result of enhanced structural interactions compared to PEG solutions or PEG + cellulose solutions. In the *q*-space, the structure factor of PEG-bentonite also shows a low *q* upturn which implies the presence of inhomogeneities, arising due to the presence of bentonite aggregate (Figure 8F).

## 4. Conclusions

By studying structuring and oscillatory shear response of polyethylene glycol (PEG) solutions in the absence and presence of small amounts of NaCMC and bentonite clay, we have explored the routes to departure from the Rouse segmental dynamics in semidilute and concentrated polyethylene glycol (PEG) solution in water, a good solvent. Concentration-dependent microstructure formation in the polymer matrix due to polymer-solvent and polymer-polymer interactions, depletion and adsorption due to presence of nanoclay, screened Coulomb interactions due to the presence of a cellulose derivative polyelectrolyte and their effect on polymer dynamics is studied with oscillatory rheology at the macro-scale. In addition, diffusion dynamics of a fluorescent probe captures mechanical properties at the micro-scale. FRAP data reveal presence of microstructures in PEG solutions. Although there are microstructures, the micro-scale diffusion coefficients were found to follow the de Gennes scaling D~$c^{-0.54}$. Optical tweezers micro-rheology has also revealed that even if there are strain induced local inhomogeneities in entangled polymer solutions, stress relaxation is Rouse-like at the microscopic scale[64]. Our studies show that the scaling beyond the elastic plateau in PEG solutions follows the Rouse, or $\omega^{1/2}$ scaling beyond a frequency $\tau_d^{-1}$. This time scale $\tau_d$ arises as a result of reversible inter-molecular bonding and dissociation kinetics, leading to a scaling exponent greater than the Rouse value of 0.5 immediately after the elastic plateau, a behaviour found in sticky associative polymers[10,13]. The range of frequencies within which this higher scaling exponent is observed is concentration dependent, shortens as the concentration is increased, until pure Rouse behaviour is observed beyond the elastic plateau. This is not surprising, because increase in PEG concentration leads to increase in the time scale of association-dissociation kinetics, leading to a decrease in $\tau_d^{-1}$ with increase in concentration, and its subsequent merging with $\tau_e^{-1}$, the frequency demarcating the end of the elastic plateau. The Rouse behaviour is recovered by addition of a polyelectrolyte, sodium carboxy methyl cellulose (PEG+CMC) in the neutral polymer matrix, possibly due to introduction of a polyelectrolyte at concentration higher than the critical overlap concentration, which leads to screening of all electrostatic interactions, and the chains behave as pure Rouse chains. However, depletion interactions and dynamic adsorption-desorption processes due to addition of bentonite clay (PEG+B), again causes a pronounced departure from the Rouse scaling at high frequencies. Addition of the clay bentonite to PEG matrix enhances the shear modulus, but this is clay-concentration limited, and yielding commences at low values of strain amplitude. Static microstructural studies from phase contrast microscopy images reveal that increasing the concentration of clay particles in PEG matrix beyond 0.1 wt% promotes clay aggregation due to depletion interactions. Micro-scale analysis of diffusion time using fluorescence recovery after photobleaching reveals heterogeneities in PEG+B, which is corroborated by the structure factor and 2D autocorrelation results. All these factors should be considered while designing PEG-based scaffolds, nanomedicine or other PEG-based biomedical products.

## Conflicts of interest

"There are no conflicts to declare".

## Acknowledgements

The authors thank the Science and Engineering Research Board (SERB), New Delhi, for funding (EMR/2016/003910), to Dr. Feroz Musthafa, CCAMP, Bangalore, for confocal microscopy experiments and Dr. K. Prasad of CSMCRI, Bhavnagar, for Rheology experiments.

## References

1 M. Goktas, G. Cinar, I. Orujalipoor, S. Ide, A. B. Tekinay and M. O. Guler, *Biomacromolecules*, 2015, **16**, 1247–1258.
2 C.-C. Lin and K. S. Anseth, *Pharm Res*, 2009, **26**, 631–643.
3 S. Senapati, A. K. Mahanta, S. Kumar and P. Maiti, *Sig Transduct Target Ther*, 2018, **3**, 7.
4 I. Ekladious, Y. L. Colson and M. W. Grinstaff, *Nat Rev Drug Discov*, 2019, **18**, 273–294.
5 S. Xin, D. Chimene, J. E. Garza, A. K. Gaharwar and D. L. Alge, *Biomater. Sci.*, 2019, **7**, 1179–1187.
6 N. S. V. Capanema, A. A. P. Mansur, A. C. de Jesus, S. M. Carvalho, L. C. de Oliveira and H. S. Mansur, *International Journal of Biological Macromolecules*, 2018, **106**, 1218–1234.
7 C.-W. Chang, A. van Spreeuwel, C. Zhang and S. Varghese, *Soft Matter*, 2010, **6**, 5157.

8  S. Lüsse and K. Arnold, *Macromolecules*, 1996, **29**, 4251–4257.
9  A. S. Muresan, J. L. A. Dubbeldam, H. Kautz, M. Monkenbusch, R. P. Sijbesma, P. van der Schoot and W. H. de Jeu, *Phys. Rev. E*, 2006, **74**, 031804.
10 Z. Zhang, C. Huang, R. A. Weiss and Q. Chen, *Journal of Rheology*, 2017, **61**, 1199–1207.
11 J. Brassinne, A. Cadix, J. Wilson and E. van Ruymbeke, *Journal of Rheology*, 2017, **61**, 1123–1134.
12 M. Rubinstein and A. N. Semenov, *Macromolecules*, 2001, **34**, 1058–1068.
13 Z. Zhang, Q. Chen and R. H. Colby, *Soft Matter*, 2018, **14**, 2961–2977.
14 M. Golkaram and K. Loos, *Macromolecules*, 2019, **52**, 9427–9444.
15 L. Leibler, M. Rubinstein and R. H. Colby, *Macromolecules*, 1991, **24**, 4701–4707.
16 D. Amin and Z. Wang, *Journal of Rheology*, 2020, **64**, 581–600.
17 B. J. Gold, C. H. Hövelmann, N. Lühmann, N. K. Székely, W. Pyckhout-Hintzen, A. Wischnewski and D. Richter, *ACS Macro Lett.*, 2017, **6**, 73–77.
18 D. Amin, A. E. Likhtman and Z. Wang, *Macromolecules*, 2016, **49**, 7510–7524.
19 G. Schmidt, A. I. Nakatani and C. C. Han, *Rheologica Acta*, 2002, **41**, 45–54.
20 V. K. Daga and N. J. Wagner, *Rheol Acta*, 2006, **45**, 813–824.
21 J. R. de Bruyn, F. Pignon, E. Tsabet and A. Magnin, *Rheol Acta*, 2008, **47**, 63–73.
22 H. A. Baghdadi, H. Sardinha and S. R. Bhatia, *J. Polym. Sci. B Polym. Phys.*, 2005, **43**, 233–240.
23 H. A. Baghdadi, J. Parrella and S. R. Bhatia, *Rheol Acta*, 2008, **47**, 349–357.
24 R. H. Colby, *Rheol Acta*, 2010, **49**, 425–442.
25 C. Weis, C. Oelschlaeger, D. Dijkstra, M. Ranft and N. Willenbacher, *Sci Rep*, 2016, **6**, 33498.
26 R. Pandey and J. C. Conrad, *Phys. Rev. E*, 2016, **93**, 012610.
27 H. A. Baghdadi, H. Sardinha and S. R. Bhatia, *J. Polym. Sci. B Polym. Phys.*, 2005, **43**, 233–240.
28 D. Bonn, M. M. Denn, L. Berthier, T. Divoux and S. Manneville, *Rev. Mod. Phys.*, 2017, **89**, 035005.
29 T. Divoux, D. Tamarii, C. Barentin and S. Manneville, *Phys. Rev. Lett.*, 2010, **104**, 208301.
30 M. Laurati, G. Petekidis, N. Koumakis, F. Cardinaux, A. B. Schofield, J. M. Brader, M. Fuchs and S. U. Egelhaaf, *The Journal of Chemical Physics*, 2009, **130**, 134907.
31 M. Takeda, Takuro Matsunaga, Toshihiko Nishida, Hitoshi Endo, Tsutomu Takahashi, and Mitsuhiro Shibayama, *Macromolecules,* 2010, **43**, 7793–7799.
32 V. Trappe and D. A. Weitz, *Phys. Rev. Lett.*, 2000, **85**, 449–452.
33 J. Ruiz-Franco, F. Camerin, N. Gnan and E. Zaccarelli, *Phys. Rev. Materials*, 2020, **4**, 045601.
34 Á. González García and R. Tuinier, *Phys. Rev. E*, 2016, **94**, 062607.
35 A. R. Denton, *Phys. Rev. E*, 2004, **70**, 031404.
36 T. Eckert and E. Bartsch, *Phys. Rev. Lett.*, 2002, **89**, 125701.
37 N. Park, V. Rathee, D. L. Blair and J. C. Conrad, *Phys. Rev. Lett.*, 2019, **122**, 228003.
38 C. A. Day. et. al. Minchul Kang, *NIH Public Access*, 2012, **13**, 1589–1600.
39 M. Kang, M. Andreani and A. K. Kenworthy, *PLoS ONE*, 2015, **10**, e0127966.
40 J. Wu, N. Shekhar, P. P. Lele and T. P. Lele, *PLoS ONE*, 2012, **7**, e42854.
41 K. L. Linegar, A. E. Adeniran, A. F. Kostko and M. A. Anisimov, *Colloid J*, 2010, **72**, 279–281.
42 D. Axelrod, D. E. Koppel, J. Schlessinger, E. Elson and W. W. Webb, *Biophysical Journal*, 1976, **16**, 1055–1069.
43 D. Blumenthal, L. Goldstien, M. Edidin and L. A. Gheber, *Sci Rep*, 2015, **5**, 11655.
44 Y. Chenyakin, D. A. Ullmann, E. Evoy, L. Renbaum-Wolff, S. Kamal and A. K. Bertram, *Atmos. Chem. Phys.*, 2017, **17**, 2423–2435.
45 P. Luca, S. S. Aime and L. Cipelletti, *Soft Matter*, 2018, **15**, 213–226.
46 R. Cerbino and V. Trappe, *Phys. Rev. Lett.*, 2008, **100**, 188102.
47 R. Furukawa, J. L. Arauz-Lara and B. R. Ware, *Macromolecules*, 1991, **24**, 599–605.
48 C. G. Lopez, S. E. Rogers, R. H. Colby, P. Graham and J. T. Cabral, *J. Polym. Sci. Part B: Polym. Phys.*, 2015, **53**, 492–501.
49 J. S. Behra, J. Mattsson, O. J. Cayre, E. S. J. Robles, H. Tang and T. N. Hunter, *ACS Appl. Polym. Mater.*, 2019, **1**, 344–358.
50 K. Hyun, J. G. Nam, M. Wilhellm, K. H. Ahn and S. J. Lee, *Rheol Acta*, 2006, **45**, 239–249.
51 R. G. Larson, *Journal of Rheology*, 2005, **49**, 1–70.
52 M. Vladkov and J.-L. Barrat, *Macromol. Theory Simul.*, 2006, **15**, 252–262.
53 C. Robin, C. Lorthioir, C. Amiel, A. Fall, G. Ovarlez and C. Le Cœur, *Macromolecules*, 2017, **50**, 700–710.
54 S. Arrese-Igor, A. Alegría and J. Colmenero, *Phys. Rev. Lett.*, 2014, **113**, 078302.
55 J. T. Kalathi, S. K. Kumar, M. Rubinstein and G. S. Grest, *Soft Matter*, 2015, **11**, 4123–4132.
56 R. Prabhakar, C. Sasmal, D. A. Nguyen, T. Sridhar and J. R. Prakash, *Phys. Rev. Fluids*, 2017, **2**, 011301.
57 R. Prabhakar, S. Gadkari, T. Gopesh and M. J. Shaw, *Journal of Rheology*, 2016, **60**, 345–366.
58 H. Lee, R. M. Venable, A. D. MacKerell and R. W. Pastor, *Biophysical Journal*, 2008, **95**, 1590–1599.
59 S. A. Oelmeier, F. Dismer and J. Hubbuch, *BMC Biophys*, 2012, **5**, 14.
60 C. M. Roland, L. A. Archer, P. H. Mott and J. Sanchez-Reyes, *Journal of Rheology*, 2004, **48**, 395–403.
61 C. G. Lopez, S. E. Rogers, R. H. Colby, P. Graham and J. T. Cabral, *J. Polym. Sci. Part B: Polym. Phys.*, 2015, **53**, 492–501.
62 H. E. King, S. T. Milner, M. Y. Lin, J. P. Singh and T. G. Mason, *Phys. Rev. E*, 2007, **75**, 021403.
63 X. Zhao, K. Urano and S. Ogasawara, *Colloid & Polymer Sci*, 1989, **267**, 899–906.
64 M. Khan, K. Regan and R. M. Robertson-Anderson, *Phys. Rev. Lett.*, 2019, **123**, 038001.